\documentclass[10pt,aps,prc,twocolumn,superscriptaddress,floatfix]{revtex4-1}
\usepackage{graphicx}
\usepackage{amsmath}
\usepackage{braket}
\usepackage{url}
\usepackage{ulem}
\usepackage{multirow}
\usepackage{amssymb} 
\usepackage{booktabs}
\usepackage{rotating}
\usepackage{bm}
\usepackage{bbm}
\usepackage{longtable}
\usepackage{subfigure}
\usepackage{tikz}
\usepackage{xcolor}
\usepackage{soul}
\usetikzlibrary{matrix}

\usepackage[colorlinks,
linkcolor=blue,
anchorcolor=blue,
urlcolor=blue,
citecolor=red]{hyperref}

\usepackage{cancel}

\begin{document}

%\title{\textit{Ab initio} study of low-lying states and spectroscopic factors in $^{48}$K and neighboring $N=28$ isotones}

\title{\textit{Ab initio} study of spectroscopic factors in $^{48}$K and neighboring $N=28$ isotones}

\author{P. Y. Wang}
\affiliation{Advanced Energy Science and Technology Guangdong Laboratory, 516007, Huizhou, China}
\affiliation{Institute of Modern Physics, Chinese Academy of Sciences, Lanzhou 730000, China}

\author{M. R. Xie}\email[]{xiemengran@impcas.ac.cn}
\affiliation{Heavy Ion Science and Technology Key Laboratory, Institute of Modern Physics, Chinese Academy of Sciences, Lanzhou 730000, China}
\affiliation{School of Nuclear Science and Technology, University of Chinese Academy of Sciences, Beijing 100049, China}

\author{Q. Yuan}
\affiliation{Heavy Ion Science and Technology Key Laboratory, Institute of Modern Physics, Chinese Academy of Sciences, Lanzhou 730000, China}
\affiliation{School of Nuclear Science and Technology, University of Chinese Academy of Sciences, Beijing 100049, China} 

%\author{H. H. Li}\email[]{lihonghuiaep@163.com}
%\affiliation{Center for Strategic Studies, China Academy of Engineering Physics, Beijing 100088, China}

\author{W. J. Huang}
\affiliation{Advanced Energy Science and Technology Guangdong Laboratory, 516007, Huizhou, China}

\author{J. G. Li}
\affiliation{Heavy Ion Science and Technology Key Laboratory, Institute of Modern Physics, Chinese Academy of Sciences, Lanzhou 730000, China}
\affiliation{School of Nuclear Science and Technology, University of Chinese Academy of Sciences, Beijing 100049, China}
\affiliation{Southern Center for Nuclear-Science Theory (SCNT), Institute of Modern Physics, Chinese Academy of Sciences, Huizhou 516000, China}

\date{\today}

\begin{abstract}

A recent \(^{47}\text{K}(d,p\gamma)^{48}\text{K}\) transfer reaction measurement has identified new excited states in \(^{48}\text{K}\) and extracted the corresponding spectroscopic factors (SFs)[\href{https://journals.aps.org/prl/abstract/10.1103/PhysRevLett.134.162504}{C. J. Paxman, \textit{et al.} PhysRevLett.134.162504 (2025)}], but they exposed sizeable discrepancies with large-scale shell-model (LSSM) calculations—especially for the low-lying states—suggesting shortcomings in the proton-neutron interaction employed by the LSSM. 
In this work, we revisit the low-lying states and SFs of \(^{48}\text{K}\) using the \textit{ab initio} valence-space in-medium similarity renormalization group (VS-IMSRG) approach based on the chiral two- and three-nucleon forces.
The calculated excitation energies reproduce the experimental data for \(^{48}\text{K}\), whereas computed SFs systematically exceed experimental values. We trace this overestimation to missing reduction factors that account for non-idealities of the transfer reaction. After introducing a phenomenological reduction factor, our VS-IMSRG results and the LSSM calculations achieve agreement with experiment. We also perform the same analysis for the neutron SFs of 
$^{47}$Ar.
Furthermore, we extend the \textit{ab initio} calculations across the $N=28$ isotones, computing excitation energies and single-neutron transfer SFs from $N=29$ isotones ranging from $^{48}$K to $^{45}$S.
By systematically removing protons from \(^{48}\text{K}\) to \(^{45}\text{S}\), we trace the evolution of the \(N=28\) shell strength via theoretical SFs values.
Our results provide a microscopic pathway to quantify the weakening of the \(N=28\) shell closure.

\end{abstract}

\pacs{}
%\keywords{}

\maketitle

\section{Introduction}

%\textit{Introduction.---}
The exploration of exotic nuclei, particularly those near the nuclear dripline, has revealed intriguing and often unexpected phenomena in nuclear structure~\cite{NOWACKI2021103866, SORLIN2008602, RevModPhys.92.015002}. Exotic nuclei, characterized by their extreme neutron-to-proton ratios, exhibit properties that markedly differ from those of stable nuclei, including altered shell structures~\cite{SORLIN2008602, RevModPhys.92.015002, PhysRevLett.96.032502, PhysRevLett.103.152503, PhysRevLett.105.102501, PhysRevC.92.034316, PhysRevLett.96.012501, Wienholtz2013, Steppenbeck2013}, halo formation~\cite{PhysRevC.109.L061304, PhysRevLett.100.192502, PhysRevLett.83.496}, and unconventional nuclear shapes~\cite{PhysRevC.89.041303, PhysRevC.105.014309, YUAN2024138331}. These exotic systems provide a stringent arena to test the interplay of nuclear forces, many-body correlations, and collectiveity, thereby pushing the boundaries of shell model (SM) frameworks and refining our understanding of effective interactions in finite nuclei.

In recent years, the evolution of shell closures near \(N=28\) has garnered considerable interest in the neutron-rich region away from stability~\cite{PhysRevLett.122.052501, PhysRevLett.99.022503, SORLIN2008602, RevModPhys.92.015002, PhysRevLett.122.222501, LI2025139609, 10.1088/1674-1137/add5dd, Xie_2024, PhysRevC.109.L041301}. 
A compact but revealing case is the doubly magic nucleus \(^{48}\mathrm{Ca}\), where the {proton \(0d_{3/2}\) and \(1s_{1/2}\) orbits} lie unusually close in energy, making the system highly sensitive to proton–neutron interaction and cross-shell neutron excitations across the \(N=28\) subshell~\cite{Xie_2024, Sun_2024}. Empirically, the weakening of the \(N=28\) closure at lower \(Z\) manifests through reduced \(2_1^+\) energies, enhanced \(B(E2)\) strengths, and the onset of deformation in neutron-rich isotopes~\cite{PhysRevLett.122.222501, PhysRevLett.122.052501, PhysRevLett.109.182501, PhysRevLett.99.022503}. Against this backdrop, the neighboring K isotopes \(^{47,48}\mathrm{K}\) (\(Z=19\)) provide a particularly clean probe, their low-lying states, level ordering, electromagnetic moments, and spectroscopic factors (SFs) directly track the spacing, mixing, and possible inversion of the {\(\pi 0d_{3/2}\) and \(\pi 1s_{1/2}\)} configurations as neutrons fill and vacate orbits around $N=28$.

Among experimental tools, transfer reactions are exceptionally valuable. By selectively populating specific nuclear states, they serve as sensitive probes of single-particle structure, enabling the extraction of SFs that quantify the distribution of single-particle strength. 
A recent study on $^{47}$K$(d,p\gamma)^{48}$K transfer reaction identified new excited states in $^{48}$K and extracted their SFs. Compared with large-scale shell-model (LSSM) calculations, the theoretical SFs are systematically larger than the experimental values, which were attributed to deficiencies in the interaction~\cite{PhysRevLett.134.162504}. 
However, such a direct attribution is not unique: SFs are not directly observable and carry an intrinsic model dependence. The theoretical SFs calculated in LSSM depended on the limited model space, with missing correlations tending to cause systematic overestimations of SFs~\cite{PhysRevLett.103.202502, BROWN2001517, LAPIKAS1993297}.
Early $(e,e'p)$ experiments showed that single-particle strengths, the so-called SFs, in stable nuclei are quenched by roughly 40\% compared to independent‑particle SM calculations~\cite{KRAMER2001267, LAPIKAS1993297, RevModPhys.69.981, JIANG2025139789}. This quenching effect has been consistently confirmed in knockout and transfer reaction studies~\cite{PhysRevLett.129.152501, PhysRevC.90.057602, PhysRevLett.131.212503, PhysRevLett.104.112701, PhysRevLett.107.032501, PhysRevLett.110.122503, PhysRevC.73.044608}. 
These observations underscore the importance of accounting for the quenching factor before concluding specific interactions. On these grounds, the conclusions in Ref.~\cite{PhysRevLett.134.162504} remain open to re-evaluation, motivating a revisit of the $^{48}$K that explicitly incorporates the SF reduction factor.

For the region below $^{48}$Ca (specifically nuclei with $N > 20$ and $Z < 20$), LSSM calculations employing the phenomenological interactions SDPF-MU ~\cite{PhysRevLett.104.012501, PhysRevC.86.051301} and SDPF-U~\cite{PhysRevC.79.014310} have been widely used. However, these interactions can yield notably different predictions for key properties, such as the low-lying spectra in $^{45}$S~\cite{PhysRevC.109.L041301}, highlighting significant sensitivities to cross-shell excitations and the proton-neutron interaction. Over recent decades, \textit{ab initio} approaches,  particularly the valence-space in-medium similarity renormalization group (VS-IMSRG)~\cite{BARRETT2013131, Hagen_2014, RevModPhys.87.1067, annurev:/content/journals/10.1146/annurev-nucl-101917-021120,  HERGERT2016165, Hu2022, PhysRevLett.118.032502}, have emerged as powerful tools for nuclear structure studies. This progress has been driven by advances in chiral effective field theory~\cite{RevModPhys.81.1773, MACHLEIDT20111}. 
The VS-IMSRG has been successfully applied to investigate the properties of nuclei with $N > 20$ and $Z < 20$.
Applications include revealing potential neutron halos in intermediate-mass nuclei through SFs and two-nucleon amplitudes~\cite{PhysRevC.109.L061304}, investigating configuration-coexisting states in $^{43,45}$S~\cite{PhysRevC.109.L041301}, and exploring shell evolution in neutron-rich P, Cl, and K isotopes~\cite{Xie_2024}.

%\textcolor{red}{In these calculations, the interactions are constrained primarily by nucleon-nucleon scattering and few-body data, reducing reliance on phenomenological fits in medium-mass nuclei while retaining controlled uncertainty estimates.}

In this work, we employ \textit{ab initio} VS-IMSRG calculations based on chiral two-nucleon (\textit{NN}) and three-nucleon ($3N$) forces to revisit the low-lying states and the SFs of $^{48}$K. Furthermore, a reduction factor is introduced to interpret new experimental SFs data obtained from a recent $^{47}$K$(d,p\gamma)^{48}$K transfer reaction. In addition, we compute and discuss the energy spectra and neutron SFs of $^{47}$Ar, confronting the VS-IMSRG and LSSM results with available data.
Additionally, the low-lying states and SFs of $N=28$ isotones below $^{48}$Ca are investigated. This analysis provides a microscopic pathway to quantify the weakening of the $N=28$ shell closure and offers new insights into shell evolution near the neutron magic number $N=28$.

\section{Method}

%\textit{Method.---}
The VS-IMSRG begins with realistic nuclear forces derived from chiral effective field theory. The bare interactions are softened and transformed to decouple a valence-space model from the rest of the Hilbert space, effectively incorporating many-body correlations. 
The continuous similarity transformation is governed by the flow equation~\cite{ annurev:/content/journals/10.1146/annurev-nucl-101917-021120, HERGERT2016165, PhysRevLett.118.032502}
\begin{equation}
    \frac{dH(s)}{ds} = [\eta(s), H(s)],
\end{equation}
where \(H(s)\) is the Hamiltonian at flow parameter \(s\), and \(\eta(s)\) is the generator of the transformation.
At sufficiently large \(s\), the transformation yields a renormalized valence-space Hamiltonian that can be directly used in LSSM calculations to obtain energies, SFs, and transition rates. Notably, the bare SF operator is used to compute the SFs for the nuclei studied in the present work.

In this work, we employ the chiral EM1.8/2.0~\cite{PhysRevC.83.031301, PhysRevC.93.011302} and the $\chi$EFT \textit{NN} N$^3$LO + 3\textit{N}(lnl) interaction~\cite{PhysRevC.101.014318} to perform the VS-IMSRG calculations, taking harmonic-oscillator basis at $\hbar\omega=16$ MeV with $e_{\rm max}=2n+l=14$ and $E_{3\rm max}=14$. The EM1.8/2.0 interaction is composed of a next-to-next-to-next-to-leading order (N$^3$LO) $NN$ interaction softened by the similarity renormalization group (SRG) evolution with momentum resolution scale $\lambda = 1.8\ {\rm fm}^{-1}$ and a next-to-next-to-leading order (N$^2$LO) $3N$ force with momentum cutoff $\Lambda= 2.0\ {\rm fm}^{-1}$. 
For the  \textit{NN} N$^3$LO + 3\textit{N}(lnl) interaction, a large SRG scale of $\lambda = 2.6\ {\rm fm}^{-1}$ for the N$^3$LO $NN$ interaction is adopted without including the induced $3N$ force. 
For the $3N$ part of the  \textit{NN} N$^3$LO + 3\textit{N}(lnl) interaction, the bare interaction is adopted in the real calculations.

The Magnus formulation of the VS-IMSRG~\cite{PhysRevC.92.034331, HERGERT2016165} is adopted to construct the effective valence space Hamiltonian, in which the $^{28}$O is taken as an inner core, and valence protons and neutrons are limited in the full $sd$- and $pf$-shell, respectively.
The ensemble normal-ordering technique detailed in Ref.~\cite{PhysRevLett.118.032502} is used in our calculations, and the VS-IMSRG code of Ref.~\cite{imsrg_code} is utilized for that matter. In practical calculations, the Magnus formalism is used with all operators truncated at the two-body level~\cite{PhysRevC.92.034331}.
Finally, the obtained effective valence space Hamiltonian is diagonalized using the \textsc{kshell} code~\cite{SHIMIZU2019372}, and the low-lying spectra and SFs are deduced.

\section{Result}

%Intuitively, the ground state of $^{48}$K only has one configuration, $\pi s^1_{1/2}d^4_{3/2}$, because of the $^{47}$K($d,\ p$) reaction, which simply populates one neutron above the $^{47}$K, meaning that the $\pi s^1_{1/2}d^4_{3/2}$ is the major proton configuration in $^{48}$K. 

\begin{figure}[h]
    \centering
    \includegraphics[width=1.0\columnwidth]{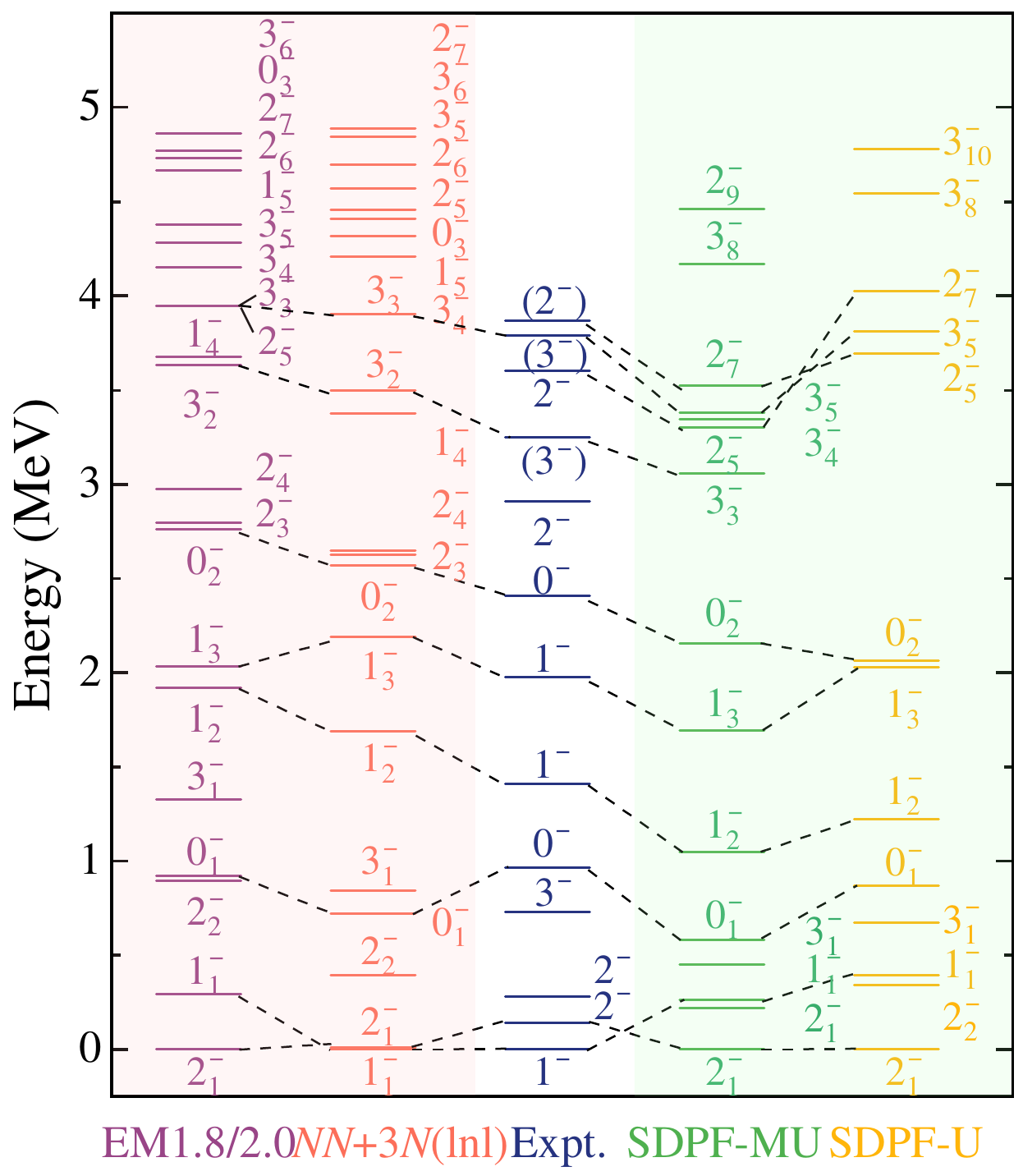}
    \caption{Spectra of $^{48}$K calculated by using VS-IMSRG with two chiral effective interactions, \textit{NN} N$^3$LO + 3\textit{N}(lnl) and EM1.8/2.0. The results predicted by LSSM using the phenomenological interactions SDPF-MU and SDPF-U are presented~\cite{PhysRevLett.134.162504}, as well as experimental data~\cite{PhysRevLett.134.162504}. }
    \label{fig:spectra_48K}
\end{figure}

%\textit{Result.---}
The excitation spectra of odd-odd nuclei provide a sensitive probe of the underlying nuclear interaction~\cite{HU2020135206, PhysRevLett.110.082502, PhysRevC.96.054305, Xie_2024, Sun_2024, PhysRevC.109.L041301}. To explore this sensitivity, we compute the excitation spectrum of $^{48}$K using the VS-IMSRG with two chiral interactions: EM1.8/2.0 and  \textit{NN} N$^3$LO + 3\textit{N}(lnl).
The results are presented in Fig.~\ref{fig:spectra_48K}. For comparison, we also include results from LSSM calculations using two phenomenological interactions: SDPF-MU and SDPF-U,  alongside the experimental data~\cite{PhysRevLett.134.162504}. Besides, for the two phenomenological interactions, we adopt the 0$\hbar\omega$ truncation. The valence space used in SDPF-U and SDPF-MU interactions is the same as in the \textit{ab initio} VS-IMSRG, namely the proton $sd$ shell and the neutron $pf$ shell.
Although entirely different procedures were used to derive SDPF-MU and SDPF-U interactions, the resulting spectra show remarkable agreement with each other.
%Although the SDPF-MU and SDPF-U interactions are derived by entirely different procedures, the resulting spectra are in close agreement. 
However, both interactions predict an inversion between the $1^-$ ground state and the first $2^-$ state, in contrast to experimental observations. Aside from this inversion, the ordering of the remaining low-lying states in $^{48}$K is well-reproduced, although the energies of the higher-lying excited states are generally underestimated.

VS-IMSRG calculations with both the \textit{NN} N$^3$LO + 3\textit{N}(lnl) and EM1.8/2.0 chiral interactions have accurately reproduce the low-lying spectrum of $^{47}$K, notably the $1/2^+$ ground state~\cite{Xie_2024}.
Specifically, for the two lowest excited states of $^{47}$K, $3/2^+$ (0.36 MeV) and $5/2^+$ (3.34 MeV), the VS-IMSRG calculations with N$^3$LO + 3\textit{N}(lnl) (EM1.8/2.0) interaction yield excitation energies of 0.60 (1.04) MeV and 3.65 (3.75) MeV, respectively. An inspection of the calculated ground state wave functions shows that the proton configuration $\pi(0d_{3/2})^4(1s_{1/2})^1$ is dominant but not unique, with $\pi(0d_{3/2})^3(1s_{1/2})^2$ configuration also contributing at a non-negligible level. A similar pattern is found in LSSM calculations employing the SDPF-MU and SDPF-U interactions.
In $^{49}$Ca, the method likewise reproduces the experimental ordering $3/2^-$ (0.00 MeV), $1/2^-$ (2.02 MeV), $7/2^-$ (3.36 MeV), and $5/2^-$ (3.99 MeV)~\cite{PhysRevC.93.031601}, with calculated energies 0.00 (0.00) MeV, 2.49 (2.23) MeV,  4.40 (4.08) MeV, and 5.29 (4.43) MeV, respectively. For $^{48}$K, the two interactions yield distinct level orderings. Only the \textit{NN} N$^3$LO + 3\textit{N}(lnl) interaction correctly predicts the $1^-$ ground state and the low-lying spectrum, whereas EM1.8/2.0—despite its success in describing ground state energies in lighter and medium-mass nuclei~\cite{PhysRevLett.126.022501, PhysRevC.105.014302}—exhibits the same inversion of the $1^-_1$ and $2^-_1$ states as phenomenological interactions and systematically overestimates excitation energies. Besides, the excitation energies of $2^-_2$ for the VS-IMSRG results with both the \textit{NN} N$^3$LO + 3\textit{N}(lnl) and the EM1.8/2.0 interactions are higher than the experimental data. This reason can be traced back to the calculated spectra of $^{49}$Ca, the VS-IMSRG results with these two interactions reproduce higher excitation energies for the $7/2^-_1$ state.

\begin{table*}[]
    \centering
    \renewcommand{\arraystretch}{1.5}
    \caption{Configuration and the calculated probabilities of the low-lying states in $^{48}$K for above $Z = 8$ and $N =20$ closure using VS-IMSRG with $NN+3N$(lnl) and EM1.8/2.0 interactions, as well as LSSM calculations using SDPF-MU and SDPF-U interaction.}
    \begin{tabular}{cccccc}
        \hline\hline
        \multirow{2}{*}{States} & \multirow{2}{*}{Configuration} & \multicolumn{4}{c}{Probability} \\
        \cline{3-6}
         & & \textit{NN}+3\textit{N}(lnl)& EM1.8/2.0& SDPF-MU & SDPF-U  \\
        \hline
        $1^-_1$ & $\pi\{(0d_{5/2})^6(0d_{3/2})^3(1s_{1/2})^2\}\otimes\nu\{(0f_{7/2})^8(1p_{3/2})^1\}$ & 38\% & 27\% & 35\% & 52\% \\
        & $\pi\{(0d_{5/2})^6(0d_{3/2})^4(1s_{1/2})^1\}\otimes\nu\{(0f_{7/2})^8(1p_{3/2})^1\}$ & 23\% & 32\% & 28\% & 14\% \\
        & $\pi\{(0d_{5/2})^6(0d_{3/2})^3(1s_{1/2})^2\}\otimes\nu\{(0f_{7/2})^7(1p_{3/2})^2\}$ & 16\% & 15\% & 17\% & 11\% \\
        $2^-_1$ & $\pi\{(0d_{5/2})^6(0d_{3/2})^4(1s_{1/2})^1\}\otimes\nu\{(0f_{7/2})^8(1p_{3/2})^1\}$ & 60\% & 64\% & 66\% & 69\% \\
        & $\pi\{(0d_{5/2})^6(0d_{3/2})^3(1s_{1/2})^2\}\otimes\nu\{(0f_{7/2})^7(1p_{3/2})^2\}$ & 19\% & 14\% & 18\% & 11\% \\
        $2^-_2$ & $\pi\{(0d_{5/2})^6(0d_{3/2})^3(1s_{1/2})^2\}\otimes\nu\{(0f_{7/2})^7(1p_{3/2})^2\}$ & 65\% & 60\% & 71\% & 70\% \\
        $0^-_1$ & $\pi\{(0d_{5/2})^6(0d_{3/2})^3(1s_{1/2})^2\}\otimes\nu\{(0f_{7/2})^8(1p_{3/2})^1\}$ & 49\% & 23\% & 52\% & 51\% \\
        & $\pi\{(0d_{5/2})^6(0d_{3/2})^4(1s_{1/2})^1\}\otimes\nu\{(0f_{7/2})^8(1p_{1/2})^1\}$ & 26\% & 47\% & 25\% & 28\% \\
        & $\pi\{(0d_{5/2})^6(0d_{3/2})^4(1s_{1/2})^1\}\otimes\nu\{(0f_{7/2})^7(1p_{3/2})^1(1p_{1/2})^1\}$ & 7.7\% & 11\% & 9.8\% & 5.6\% \\
        $3^-_1$ & $\pi\{(0d_{5/2})^6(0d_{3/2})^3(1s_{1/2})^2\}\otimes\nu\{(0f_{7/2})^8(1p_{3/2})^1\}$ & 67\% & 61\% & 73\% & 74\% \\
        $1^-_2$ &  $\pi\{(0d_{5/2})^6(0d_{3/2})^4(1s_{1/2})^1\}\otimes\nu\{(0f_{7/2})^8(1p_{3/2})^1\}$ & 35\% & 25\% & 36\% & 51\% \\
        & $\pi\{(0d_{5/2})^6(0d_{3/2})^3(1s_{1/2})^2\}\otimes\nu\{(0f_{7/2})^8(1p_{3/2})^1\}$ & 25\% & 24\% & 31\% & 17\% \\
        & $\pi\{(0d_{5/2})^6(0d_{3/2})^3(1s_{1/2})^2\}\otimes\nu\{(0f_{7/2})^8(1p_{1/2})^1\}$ & 14\% & 13\% & 13\% & 7.9\% \\
        %{\color{red}
        $1^-_3$ &  $\pi\{(0d_{5/2})^6(0d_{3/2})^4(1s_{1/2})^1\}\otimes\nu\{(0f_{7/2})^8(1p_{1/2})^1\}$ & 60\% & 54\% & 60\% & 63\% \\
        & $\pi\{(0d_{5/2})^6(0d_{3/2})^3(1s_{1/2})^2\}\otimes\nu\{(0f_{7/2})^7(1p_{3/2})^1(1p_{1/2})^1\}$ & 19\% & 13\% & 18\% & 12\% \\
        & $\pi\{(0d_{5/2})^6(0d_{3/2})^4(1s_{1/2})^1\}\otimes\nu\{(0f_{7/2})^6(1p_{3/2})^2(1p_{1/2})^1\}$ & 3.9\% & 3.4\% & 3.7\% & 3.0\% \\%}
        $0^-_2$ & $\pi\{(0d_{5/2})^6(0d_{3/2})^4(1s_{1/2})^1\}\otimes\nu\{(0f_{7/2})^8(1p_{1/2})^1\}$ & 37\% & 20\% & 41\% & 46\% \\
        & $\pi\{(0d_{5/2})^6(0d_{3/2})^3(1s_{1/2})^2\}\otimes\nu\{(0f_{7/2})^8(1p_{3/2})^1\}$ & 28\% & 49\% & 34\% & 32\% \\
        & $\pi\{(0d_{5/2})^6(0d_{3/2})^3(1s_{1/2})^2\}\otimes\nu\{(0f_{7/2})^7(1p_{3/2})^1(1p_{1/2})^1\}$ & 11\% & 3.2\% & 8.3\% & 6.5\% \\
        \hline\hline
    \end{tabular}
    \label{tab:occ}
\end{table*}

\begin{table}%[ht]
    \centering
    \caption{The comparison of the $^{48}$K SFs obtained from \textit{ab inito} VS-IMSRG using \textit{NN} N$^3$LO + 3\textit{N}(lnl) and EM1.8/2.0, labeled by $ C^2S_{1}$ and $C^2S_{2}$, respectively, and LSSM calculations using SDPF-MU and SDPF-U, labeled by $ C^2S_{3}$ and $C^2S_{4}$, respectively,  along with experimental data. The LSSM
    results %calculated by LSSM with SDPF-MU and SDPF-U interactions 
    and experimental data are taken from Ref.~\cite{PhysRevLett.134.162504}. The values of the $0f_{5/2}$ partial wave $C^2S$ are too small (with the $^*$ label), which need to multiply a factor of $10^{-3}$. }
    \renewcommand{\arraystretch}{1.5}
    \begin{tabular}{cccccccc}
        \hline
        \hline
%         $E_{\rm x}$ & $J^\pi$ & $nlj$ & $C^2S_{\rm Expt}$ & $ C^2S_{NN+3N(\rm lnl)}$ & $C^2S_{\rm EM1.8/2.0}$ & $C^2S_{\rm SDPF-MU}$ & $C^2S_{\rm SDPF-U}$ \\
        $E_{\rm x}$ & $J^\pi$ & $nlj$ & $C^2S_{\rm Expt}$ & $ C^2S_{1}$ & $C^2S_{2}$ & $C^2S_{3}$ & $C^2S_{4}$ \\
         \hline
         0.00 & 1$^-_1$ & 1$p_{3/2}$ & 0.24(5) & 0.35 & 0.45 & 0.40 & 0.21 \\
         0.143(1) & 2$^-_1$ & 1$p_{3/2}$ & 0.42(8) & 0.85 & 0.85 & 0.86 & 0.84 \\
          & & 0$f_{5/2}$ & & 2.30$^*$ & 8.50$^*$ & 0.00$^*$ & 2.30$^*$ \\
         0.279(1) & 2$^-_2$ & 1$p_{3/2}$ & $<$0.03 & 0.02 & 0.00 & 0.01 & 0.05 \\
          & & 0$f_{5/2}$ & & 4.60$^*$ & 6.40$^*$ & 2.7$^*$ & 0.30$^*$ \\
         0.728(3) & 3$^-_1$ & 0$f_{7/2}$ & $<$0.04 & 0.07 & 0.05 & 0.06 & 0.05 \\
          & & 0$f_{5/2}$ & & 4.00$^*$ & 9.40$^*$ & 3.00$^*$ & 3.40$^*$ \\
         0.967(2) & 0$^-_1$ & 1$p_{1/2}$ & 0.26(5) & 0.40 & 0.68 & 0.40 & 0.38 \\
         1.409(3) & 1$^-_2$ & 1$p_{3/2}$ & 0.24(5) & 0.36 & 0.25 & 0.35 & 0.54 \\
         1.978(4) & 1$^-_3$ & 1$p_{1/2}$ & 0.50(10) & 0.92 & 0.78 & 0.88 & 0.84 \\
         2.407(6) & 0$^-_2$ & 1$p_{1/2}$ & 0.34(7) & 0.56 & 0.26 & 0.56 & 0.58 \\
         \hline
         \hline
    \end{tabular}
    \label{tab:S_spectroscopic}
\end{table}

Table~\ref{tab:occ} lists the dominant configurations of the low-lying states
%\sout{Table~\ref{tab:occ} lists the configurations of the low-lying states}
in $^{48}$K above the $Z=8$ and $N=20$ closures and their contributions using \textit{ab initio} VS-IMSRG and LSSM with SDPF-MU and SDPF-U interactions. 
Near $^{48}$K, {$\pi d_{3/2}$ and $\pi s_{1/2}$} orbitals are near-degenerate, the structures are highly sensitive to the proton–neutron interaction and the resulting configuration mixing.
Results show that the lowest $1^-$ state is primarily {$\pi d^4_{3/2}s^1_{1/2}$ and $\pi d^3_{3/2}s^2_{1/2}$} coupled to the neutron orbital $1p_{3/2}$, although the relative weights differ among the nuclear interactions. 
In $1^-_1$ state, the contribution of the {$\pi d^3_{3/2}s^2_{1/2}$} is smaller than the $\pi d^4_{3/2}s^1_{1/2}$ with the EM1.8/2.0 interaction, in contrast to the result obtained with the \textit{NN} N$^3$LO + 3\textit{N}(lnl) interaction, as well as the LSSM calculations with SDPF-MU and SDPF-U interactions. 
Moreover, the calculations from \textit{NN} N$^3$LO + 3\textit{N}(lnl) interaction also give that the {$\pi d^4_{3/2}s^1_{1/2}$} coupled to the neutron orbital $1p_{3/2}$ is smaller than that of other calculations.
The results indicate that the \textit{NN} N$^3$LO + 3\textit{N}(lnl) interaction more faithfully captures the relevant proton–neutron correlations, even in the presence of near-degeneracy between the $1_1^-$ and $2_1^-$ states.

We next probe the role of the relevant interaction matrix elements.  
As noted in Ref.~\cite{PhysRevLett.134.162504}, the correct ordering of the $^{48}$K low-lying spectra can be reproduced by increasing the $\langle p_{3/2}s_{1/2}|V_{\text {int}}|p_{3/2}d_{3/2}\rangle_{J=2, T=0}$ and the $\langle p_{3/2}s_{1/2}|V_{\text {int}}|p_{3/2}s_{1/2}\rangle_{J=1, T=0}$ by approximately 1 MeV. 
%In the \textit{NN} N$^3$LO + 3\textit{N}(lnl) interaction, these two matrix element values are -0.28 MeV and -0.48 MeV. The second value is larger than that of SDPF-MU (-0.84 MeV) but smaller than that of SDPF-U (-0.38 MeV).
For these two matrix elements, the monopole term $\langle p_{3/2}s_{1/2}|V_{\text {int}}|p_{3/2}s_{1/2}\rangle$ can be controlled via modifying the single-particle energies, whereas the $\langle p_{3/2}s_{1/2}|V_{\text {int}}|p_{3/2}d_{3/2}\rangle$ decomposes into components with different values of $J$. 
In this work, we describe the interaction matrix elements under the proton-neutron representation. 
The matrix element of $\langle p_{3/2}s_{1/2}|V_{\text {int}}|p_{3/2}d_{3/2}\rangle_{J=2}$ in proton-neutron representation is $-0.05$ MeV in SDPF-MU and $-0.11$ MeV in SDPF-U.
In the \textit{NN} N$^3$LO + 3\textit{N}(lnl), it is $-0.28$ MeV, stronger than in the phenomenological nuclear interactions.
This enhancement is consistent with the adjustment in Ref.~\cite{PhysRevLett.134.162504}, where the $\langle p_{3/2}s_{1/2}|V_{\text {int}}|p_{3/2}d_{3/2}\rangle_{J=2, T=0}$ is enhanced by approximately 1 MeV.
However, they neglect the influence of the $J=1$ component. For the matrix element $\langle p_{3/2}s_{1/2}|V_{\text {int}}|p_{3/2}d_{3/2}\rangle_{J=1}$ in proton-neutron representation, the values are $0.11$ MeV in \textit{NN} N$^3$LO + 3\textit{N}(lnl), $0.0037$ MeV in SDPF-MU and $-0.11$ MeV in SDPF-U. This comparatively larger $J=1$ component in \textit{NN} N$^3$LO + 3\textit{N}(lnl) may also help reproduce the ground state $1^-$, so the correct ordering is obtained without modifying the matrix elements.

Moreover, we extend our VS-IMSRG calculations to the SFs of $^{48}$K for comparison. The results, together with the experimental data and LSSM calculations using both SDPF-MU and SDPF-U interactions, are compiled in Table~\ref{tab:S_spectroscopic}. 
Notably, all theoretical calculations, including VS-IMSRG, systematically overpredict the experimental SFs, in line with the well-established suppression trends observed across different reaction probes.
Moreover, the VS-IMSRG SFs obtained with the \textit{NN} N$^3$LO + 3\textit{N}(lnl) interaction show remarkable agreement with those from the LSSM using the SDPF-MU interaction.
Lepton-induced $(e,e'p)$ knockout reactions have long demonstrated that measured cross sections are only about 60\% of those predicted by LSSM SFs~\cite{SICK2007447, LAPIKAS1993297}. Proton-induced quasi-free knockout $(p,2p)$ reactions show a similar 30–40\% reduction relative to LSSM predictions~\cite{3e0e9c2d81434dde93243710c88fe298, GOMEZRAMOS2018511, HOLL2019682, PhysRevLett.130.172501}. Transfer reactions yield comparable suppression factors~\cite{PhysRevLett.110.122503}, and direct comparisons between SFs extracted from $(d,p)$ or $(p,d)$ reactions and LSSM values likewise indicate a 30–40\% deficit~\cite{PhysRevC.73.044608, PhysRevLett.104.112701, PhysRevLett.131.212503}. 
Therefore, the trends seen in Table~\ref{tab:S_spectroscopic} are fully consistent with these earlier observations from transfer reactions and lepton-induced $(e,e'p)$ knockout reactions~\cite{SICK2007447, LAPIKAS1993297}.

These systematic quenching effects—manifested both in direct reaction cross sections and in extracted SFs—highlight the critical role of many-body correlations, rather than deficiencies in the interaction. Within the LSSM framework, the nuclear many-body wavefunction is restricted to a limited valence space, neglecting higher-orbital correlations and core polarization. This restriction systematically leads to an overestimation of theoretical SFs, a trend that is also evident in the present work. Moreover, the extraction of experimental SFs is inherently model dependent. Choices regarding the reaction mechanism, underlying model assumptions, and the parametrization of optical potentials, among other inputs, introduce substantial uncertainties~\cite{TIMOFEYUK2020103738, MA201926, AUMANN2021103847, BERTULANI2006372, PhysRevLett.102.232501, Gaudefroy2006}.
Collectively, these considerations indicate that elucidating the microscopic origin of quenching requires a theoretically consistent unification of nuclear reactions and nuclear structure—ideally within an \textit{ab initio} framework that treats their interdependence on equal footing. A fully unified \textit{ab initio} treatment, however, remains beyond the current state of the art. 
Consequently, applying an empirical reduction factor becomes essential when comparing theory with experiment, as it compensates for the insufficient treatment of many-body correlations in such models.

To further analyze these low-excitation states, we used the particle addition sum-rule~\cite{brown2005lecture} with the following formula:
\begin{equation}
    \sum_{j_f}\frac{2j_f+1}{2j_i+1}C^2S(j) = (2j+1)-\langle j \rangle,
\end{equation}
here $j_i$ is the spin quantum number of the ground state of $^{47}$K, $j_f$ is the spin quantum number for various states of $^{48}$K, $C^2S(j)$ is the corresponding SF for the partial wave $j$, and $\langle j \rangle$ is the occupation number of the orbit $j$ in the ground state of $^{47}$K. Considering the low-lying states shown in Table~\ref{tab:S_spectroscopic}, we find that most of the $p$-wave strength is concentrated in these states. For the four interactions, the strength of the $p_{3/2}$ ranges from 87\% to 90\%, and the strength of the $p_{1/2}$ ranges from 83\% to 94\%. In contrast, the $f$-wave strength is much lower, with the $f_{7/2}$ ranging from 33\% to 59\%, and the $f_{5/2}$ between 0.3\% and 1.2\%. This indicates that most $f$-wave strength resides in higher-lying states beyond those listed.

%\sout{These consistent suppressions—from direct-reaction cross sections to extracted SFs—underscore the pervasive impact of many-body correlations. In the LSSM, the many-body wavefunction is constructed within a limited valence space, neglecting correlations from higher orbitals and core polarization. As a result, theoretical SFs are typically larger than experimental ones, a trend likewise evident in the present work. Both LSSM and VS-IMSRG results systematically overestimate the experimental SFs, and it has therefore become customary to apply an empirical reduction factor when comparing theory with experiment.}
%. Therefore, when comparing theoretical SFs with experimental values, it is customary to introduce a reduction factor. 

\begin{figure}[ht]
    \centering
    \includegraphics[width=1.0\columnwidth]{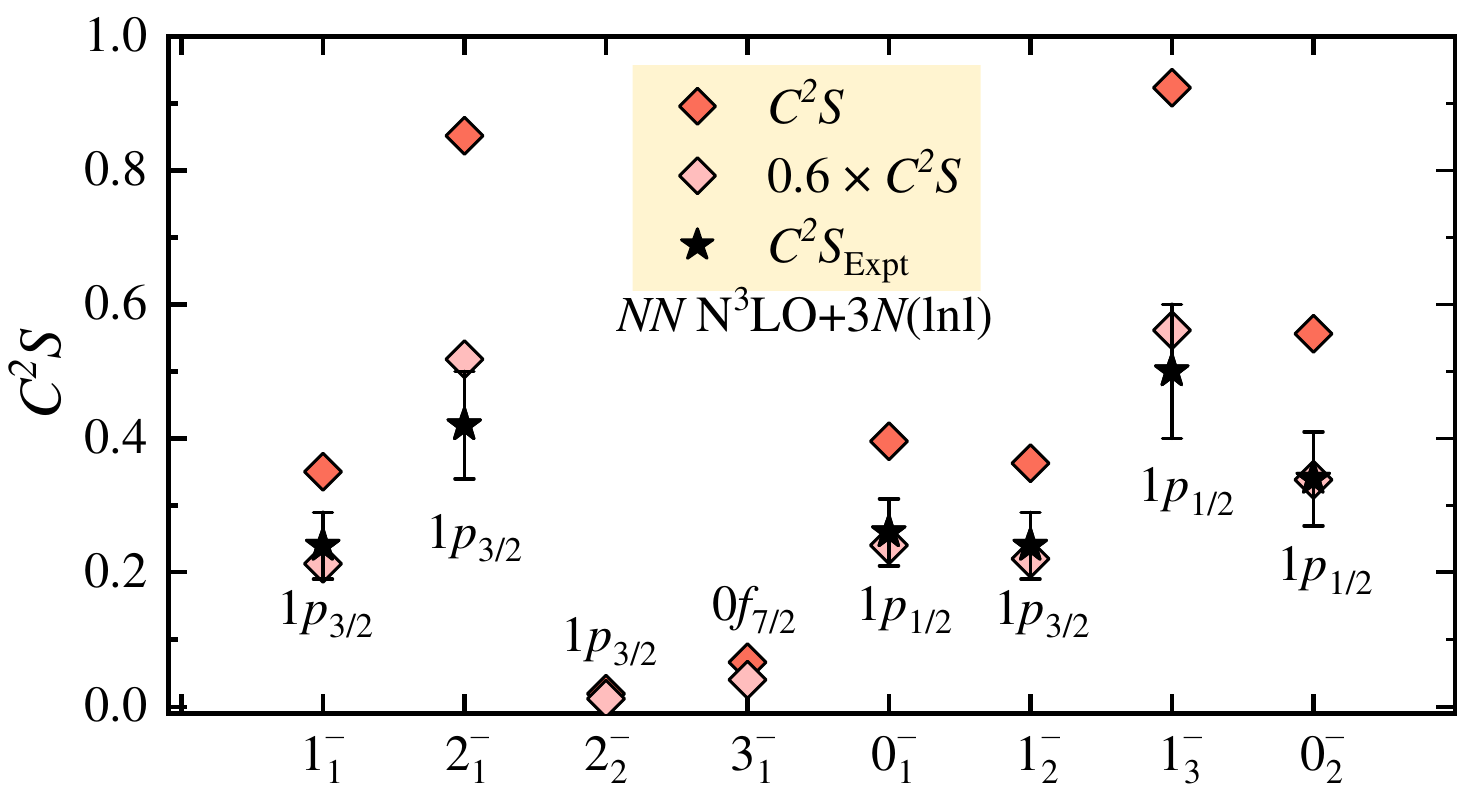}
    \includegraphics[width=1.0\columnwidth]{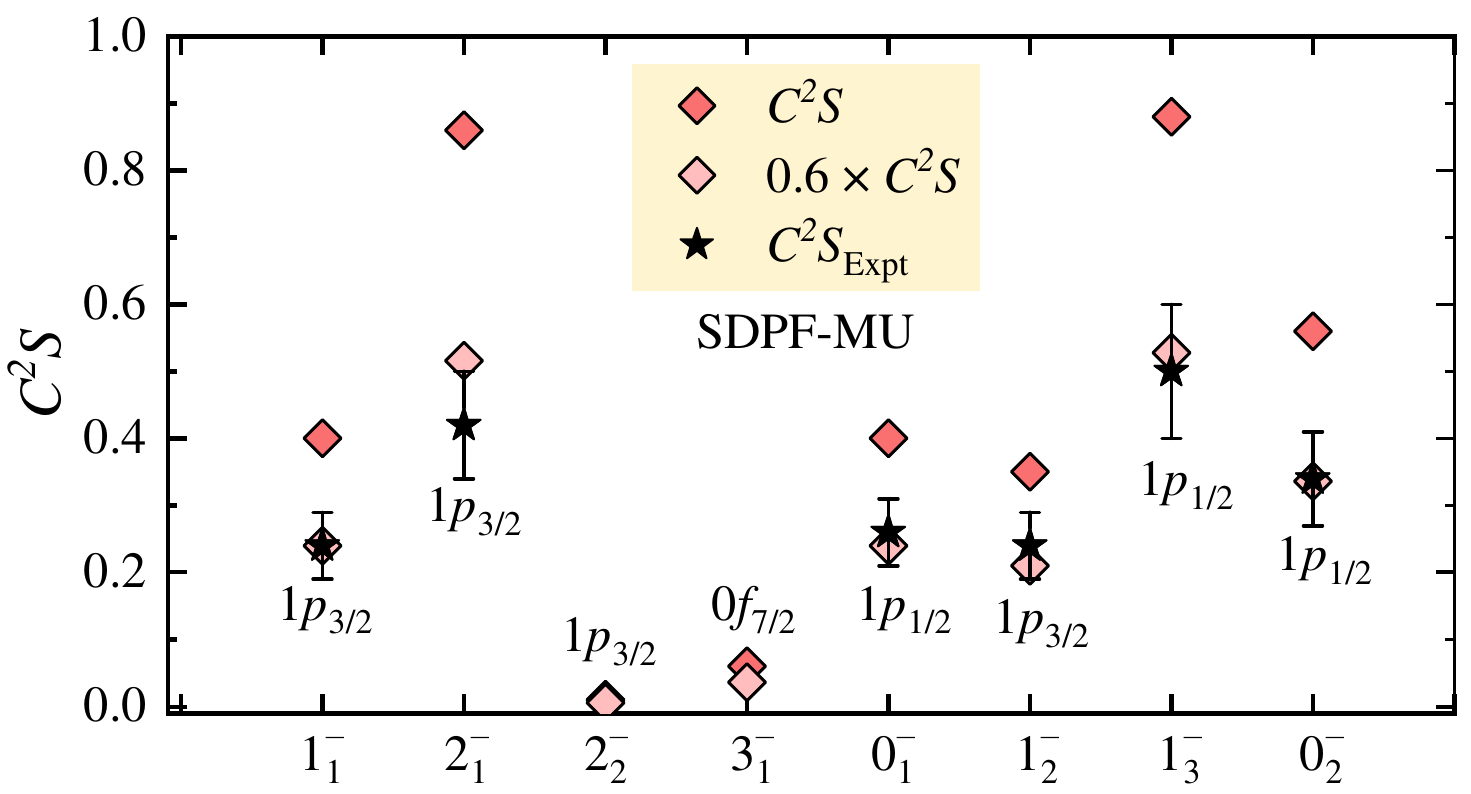}
    \caption{The $^{48}$K SFs calculated by using the \textit{NN} + 3\textit{N}(lnl) chiral interaction (above) and the SDPF-MU interaction (below) compared with the experimental results. The red diamonds represent the SFs directly calculated by the VS-IMSRG and LSSM, while the pink diamonds represent the SFs multiplied by the reduction factor. The experimental results are taken from Ref.~\cite{PhysRevLett.134.162504}. }
    \label{fig:C2S_lnl_SDPF-MU}
\end{figure}

Figure~\ref{fig:C2S_lnl_SDPF-MU} provides a further compares between experimental SFs extracted from $^{47}$K$(d,p\gamma)^{48}$K reaction and theoretical calculations. The latter include LSSM calculations with the SDPF-MU interaction and \textit{ab initio} VS-IMSRG calculations based on the \textit{NN} N$^3$LO + 3\textit{N}(lnl) interaction, presented both with and without applying a reduction factor of 0.6. Incorporating this reduction factor markedly improves the agreement with the experiment. This underscores the necessity of introducing an empirical reduction factor when comparing valence-space calculations (e.g., LSSM or VS-IMSRG-derived Hamiltonians) with experimental SFs.

%These include LSSM calculations using SDPF-MU interaction and \textit{ab initio} results based on the \textit{NN} N$^3$LO +3\textit{N}(lnl) interaction, both unscaled and scaled by a reduction factor of 0.6. A systematic overestimation persists across all theoretical SFs, particularly pronounced for the $2^-_1$ and $1^-_3$ states. After applying the 0.6 reduction factor, the theoretical SFs show significantly improved agreement with experimental data. This demonstrates that the empirical reduction factor is essential when comparing valence-space calculations (e.g., LSSM or VS-IMSRG-derived Hamiltonians) to experimental SFs.
%extracted from $^{47}$K$(d,p\gamma)^{48}$K transfer reaction
% - consistent with earlier findings. 

\begin{figure}
    \centering
    \includegraphics[width=1.0\columnwidth]{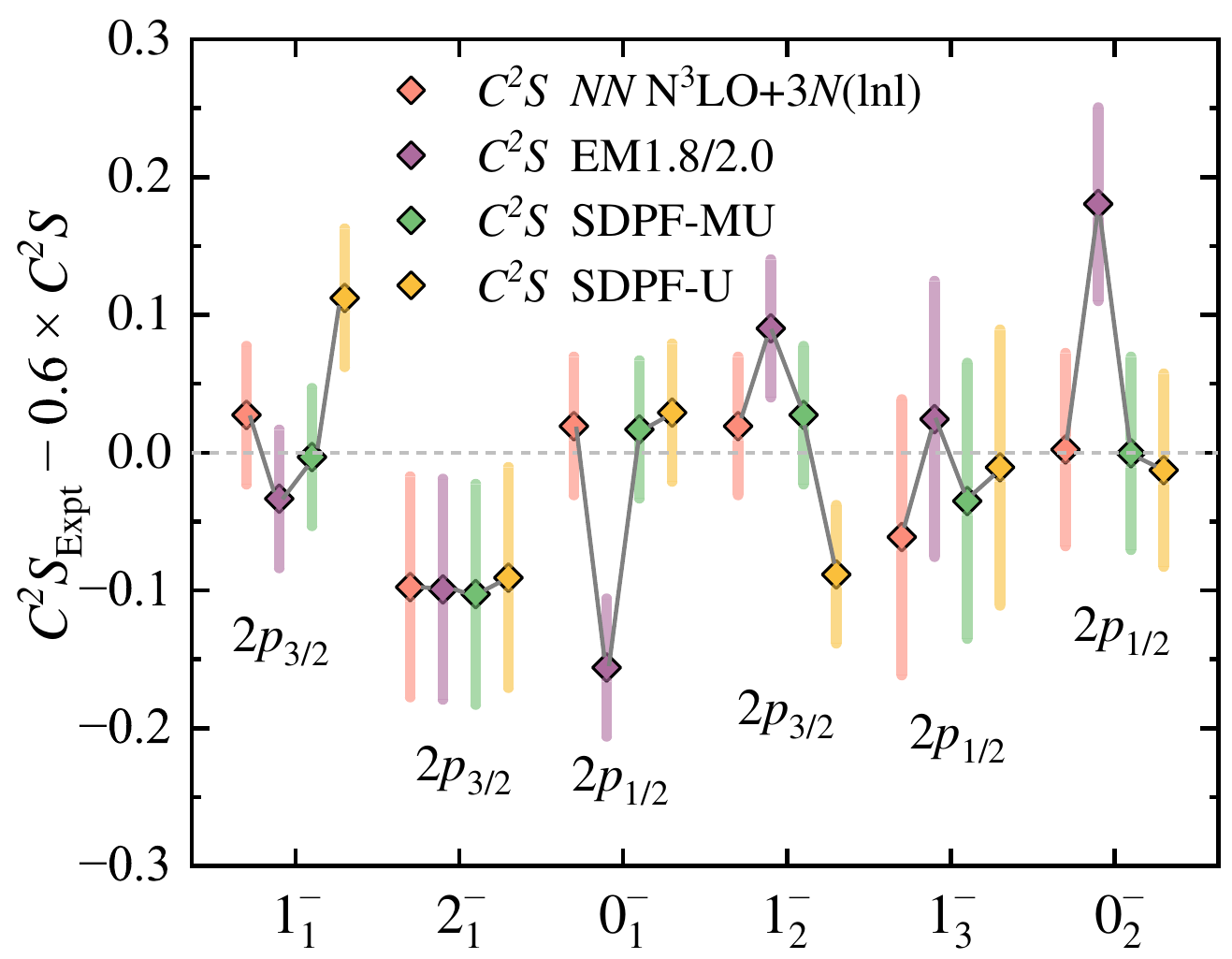}
    \caption{The difference between the experimental SFs and the calculated results multiplied by the reduction factor 0.6. }
    \label{fig:comparison}
\end{figure}

%\textbf{For states, one-by-pne to show the discrepancy of the SFs between experimental data and EM1.8/2.0  and LSSM calculations using SDPF-U interaction. More discussion for the SDPF-U should be added.}

Furthermore, we apply this reduction factor to all theoretical SFs for the low-lying states of $^{48}$K to evaluate the four interactions considered, except for the $2^-_2$ and $3^-_1$ states, for which precise experimental data are unavailable.
The deviations between experimental SFs and scaled theoretical SFs are shown in Fig.~\ref{fig:comparison}, providing a rigorous basis for evaluating the relative accuracy of theoretical models in describing the experimental SFs.
For the $1^-_1$ state, theoretical SFs show good agreement with experimental data, except for the LSSM calculation using the SDPF-U interaction. At the $2^-_1$ state, although different interactions yield slightly different level orderings, the theoretical SFs are very similar and, after applying the reduction factor, all exceed the experimental value of approximately 0.10. 
As shown in Table~\ref{tab:occ}, all four nuclear interactions show the same trend: the dominant configuration is {$\pi d^4_{3/2}s^1_{1/2}\otimes \nu p_{3/2}$} with a weight of about 60-70\%, while a subdominant configuration, {$\pi d^3_{3/2}s^2_{1/2}\otimes \nu f^7_{7/2} p^2_{3/2}$}, contributes roughly 10-20\%. This discrepancy in the $2^-_1$ state may be caused by the enhanced contribution of the $\pi d_{3/2}^4 s{1/2}^1  \otimes \nu p_{3/2}$ component in the wave functions obtained with these four interactions.
In the cases of the $0^-_1$ and $0^-_2$ states, only the VS-IMSRG calculation with EM1.8/2.0 interaction exhibits significant deviations from the experiment results. 
From Table~\ref{tab:occ}, we can see that the $0^-_1$ state is dominated by the {$\pi d^3_{3/2}s^2_{1/2} \otimes \nu p_{3/2}$} configuration for the three interactions other than EM1.8/2.0, where its probability is about twice that of {$\pi d^4_{3/2}s^1_{1/2}\otimes \nu p_{3/2}$} configuration. In contrast, this trend is reversed for EM1.8/2.0. This reversion is also observed in the $0^-_2$ state. In EM1.8/2.0 interaction, the dominant configuration is {$\pi d^4_{3/2}s^1_{1/2}$}, in clear contrast to the results obtained with the other nuclear interactions. 
A similar trend is observed in the $1^-$ states. For comparison, the VS-IMSRG with \textit{NN} N$^3$LO + 3\textit{N}(lnl) and the LSSM with SDPF-MU reproduce the experimental values accurately, while the VS-IMSRG with EM1.8/2.0 overestimates and the LSSM with SDPF-U underestimates the $1^-_1$ state, with the opposite trend for the $1^-_2$ state. As shown in Table~\ref{tab:occ}, EM1.8/2.0 consistently underestimates and SDPF-U consistently overestimates the dominant configuration in both $1^-_1$ and $1^-_2$ states, leading to larger discrepancies in these interactions.
For EM1.8/2.0 interaction, this discrepancy likely arises because the calculated $3/2^+_1$ level in $^{47}$K is overestimated by about 0.7~MeV; therefore, a reliable interpretation of the results for odd--odd nuclei requires an accurate reproduction and understanding of the simpler spectra of the neighboring odd--even nuclei.

A recent $^{47}\text{K}(d,p\gamma)^{48}\text{K}$ transfer reaction study~\cite{PhysRevLett.134.162504} compared experimental low-lying spectra and SFs with LSSM calculations using SDPF-MU and SDPF-U interaction, and attributed the discrepancies in both spectra and SFs to deficiencies in the proton–neutron interaction adopted in LSSM. However, the systematic overestimation of SF relative to experimental is not unique to LSSM. The VS-IMSRG calculations with the \textit{NN} N$^3$LO + 3\textit{N}(lnl) and EM1.8/2.0 interactions also systematically overpredict the experimental values.
%Compared to the experimental results, the SFs calculated by  VS-IMSRG using the \textit{NN} N$^3$LO + 3\textit{N}(lnl) and EM1.8/2.0 interactions also show systematic overestimation. 
The study~\cite{PhysRevLett.134.162504} in question did not adequately account for the universality of this systematic quenching, instead attributing the discrepancy mainly to the near-degeneracy of the {$\pi d_{3/2}^4s_{1/2}^1$} and {$\pi d_{3/2}^3s_{1/2}^2$} configurations, and suggesting that inaccuracies in describing the configuration mixing require adjustments to specific proton-neutron matrix elements. While such modifications partially improved the reproduction of the level structure, it remained unclear whether they also resolved the systematic overestimation of SFs. 
The \textit{ab initio} VS-IMSRG calculations with the chiral \textit{NN} N$^3$LO + 3\textit{N}(lnl) interaction correctly reproduce the ordering of the ground state ($1^-_1$) and first excited state ($2^-_1$). 
Contrary to the conclusions of Ref.~\cite{PhysRevLett.134.162504}, the deficiencies in proton-neutron interactions may not be the dominant reason for causing the overestimation of SFs. 
As discussed above, after introducing the reduction factor to correct the theoretical SFs, the theoretical results of SFs show significantly improved agreement with experimental data.
In conclusion, without the reduction factor, comparing the theoretical SFs with the experiment is unreasonable.

\begin{figure}
    \centering
    \includegraphics[width=1.0\columnwidth]{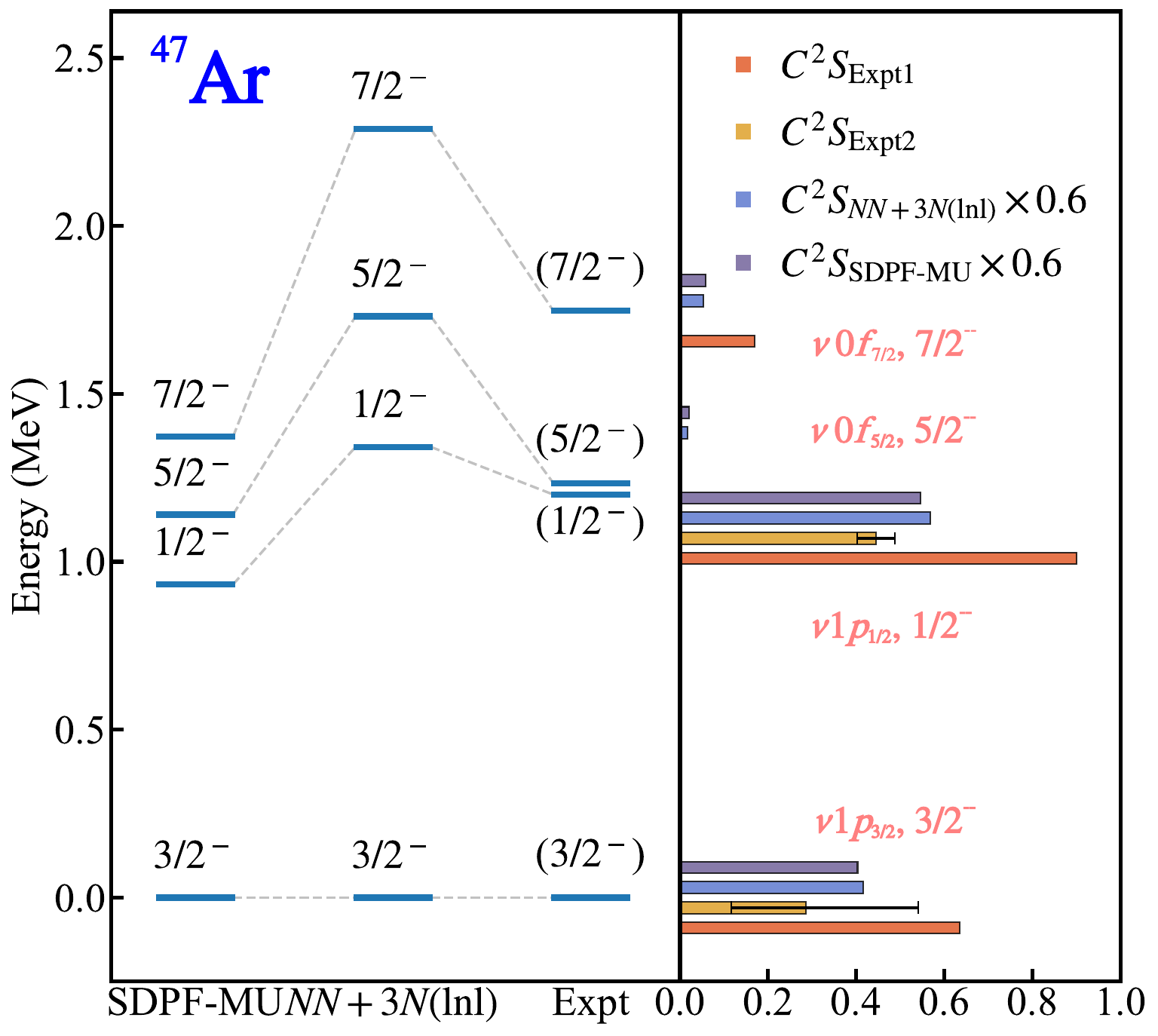}
    \caption{Low-lying spectra of $^{47}$Ar and the corresponding neutron SFs. Theoretical results are from LSSM (SDPF-MU) and VS-IMSRG (\textit{NN} N$^3$LO + 3\textit{N}(lnl)) calculations. Experimental spectra are taken from Ref.~\cite{PhysRevLett.101.032501}, and the experimental SFs (Expt1, Expt2) from Refs.~\cite{Gaudefroy2006} and~\cite{BRADT2018155}. Red annotations indicate the orbital and spin-parity associated with each SF. Theoretical SFs include quenching via an empirical reduction factor.}
    \label{fig:47Ar}
\end{figure}

Similar to $^{48}\text{K}$, the structure and SFs of its isotone $^{47}\text{Ar}$ have attracted experimental attention via the $d(^{46}\mathrm{Ar},^{47}\mathrm{Ar})p$ transfer reaction~\cite{PhysRevLett.97.092501, Gaudefroy2006} and resonant proton scattering measurements~\cite{BRADT2018155}, however, the existing studies remain relatively limited. We therefore extend the same analysis to $^{47}$Ar. Fig.~\ref{fig:47Ar} shows the low-lying spectra and neutron SFs of $^{47}$Ar computed with LSSM (SDPF-MU) and VS-IMSRG (\textit{NN} N$^3$LO + 3\textit{N}(lnl)). The SFs are scaled by the empirical reduction factor of 0.6 introduced above, and the results are compared with available experimental data.
Both approaches reproduce the ordering of the low-lying states. Overall, LSSM excitation energies lie slightly below experiment, whereas VS-IMSRG results are slightly higher~\cite{PhysRevLett.101.032501}. 
For SFs, the two theories agree closely, differing by less than 0.02 after applying the reduction factor.
In the $d(^{46}\mathrm{Ar},^{47}\mathrm{Ar})p$ reaction, the experimental SFs of the ground and first excited states agree well with the unscaled LSSM and VS-IMSRG results. Once the reduction factor is included, theory typically undershoots by about 0.1–0.3~\cite{Gaudefroy2006}.
In contrast, SFs extracted from isobaric analogue resonances in inverse kinematics~\cite{BRADT2018155} exhibit clear quenching, and the reduction factor corrected theory then falls within the experimental uncertainties.

\begin{figure*}[]
    \centering
    \includegraphics[width=2\columnwidth]{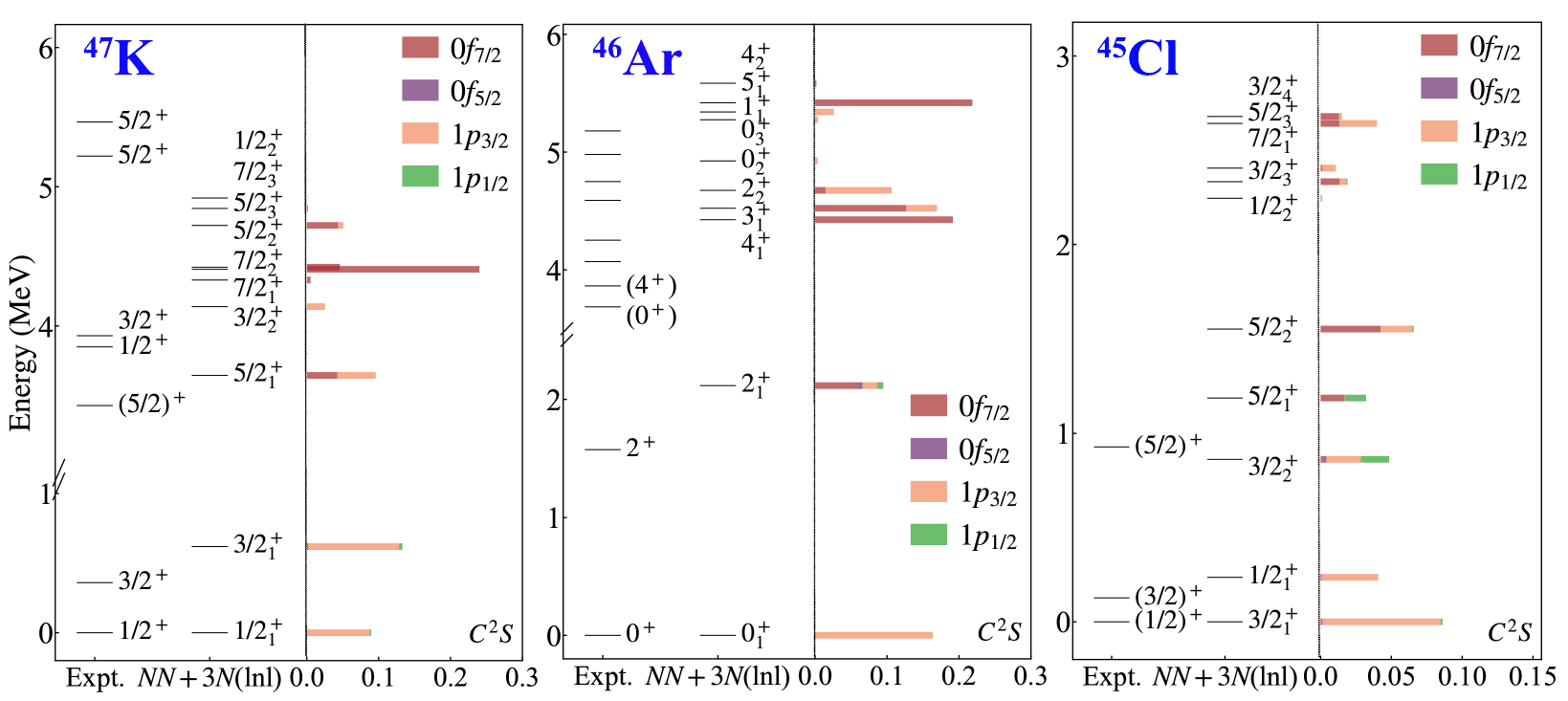}
    \caption{Low‑lying spectra of the $N=28$ isotones ($^{47}$K, $^{46}$Ar, $^{45}$Cl) and corresponding neutron SFs, calculated with VS‑IMSRG using the \textit{NN} N$^3$LO + 3\textit{N}(lnl) interaction. Experimental excitation energies for $^{47}$K, $^{46}$Ar, and $^{45}$Cl are taken from Refs.~\cite{PhysRevLett.134.162504, nndc, tghm-sszh}. SFs for the neutron partial waves $0f_{7/2}$, $0f_{5/2}$, $1p_{3/2}$, and $1p_{1/2}$ are  normalization by their degeneracy factors $(2j+1)$.
    }
    \label{fig:47K46Ar45Cl}
\end{figure*}

Using VS-IMSRG with the \textit{NN} N$^3$LO + 3\textit{N}(lnl) interaction, we have systematically investigated the $N=28$ isotonic chain in K, Ar, and Cl nuclei. Fig.~\ref{fig:47K46Ar45Cl} presents the calculated low-lying excitation spectra of $^{47}$K, $^{46}$Ar, and $^{45}$Cl. For $^{47}$K, the experimentally observed $1/2^+$ ground state arises from the inversion between the proton $1s_{1/2}$ and $0d_{3/2}$ orbitals at $N=28$~\cite{BJERREGAARD1967568}, which is well reproduced by our calculations, together with a good description of the low-lying excited states. For $^{46}$Ar, the obtained excitation energy of $0_2^+$ state is higher than that of experimental data by approximately 1 MeV, which is similar to the case of $1/2_2^+$ state of $^{47}$K.
Moreover, the VS-IMSRG calculations could provide good descriptions for the $0_2^+$ states in $^{46}$Ar isotones~\cite{YUAN2024138331}, $^{44}$S, $^{42}$Si, and $^{40}$Mg. 
For $^{45}$Cl, earlier studies~\cite{Sorlin2004} suggested a $1/2^+_1$ ground state, whereas more recent measurements~\cite{PhysRevC.109.044320} identified the ground state as $3/2^+_1$. Our VS-IMSRG results are consistent with this updated assignment.

%In particular, for the benchmark $^{47}\mathrm{K}/^{48}\mathrm{K}$ pair—where previous SF analyses focused only on the ground state—we extend the calculation to include excited states of $^{47}$K. The same approach is applied to the $^{46}$Ar/$^{47}$Ar and $^{45}$Cl/$^{46}$Cl pairs, providing a consistent framework to probe orbital occupancies and the evolution of the $N=28$ shell form $_{19}$K to $_{17}$Cl.}

\begin{table}[]
    \centering
    \renewcommand{\arraystretch}{1.5}
    \caption{The single-neutron removal SFs of the $N=29$ isotones, $^{48}$K($1_1^-$), $^{47}$Ar($3/2_1^-$), $^{46}$Cl($2_1^-$) and $^{45}$S($3/2_1^-$) calculated by VS-IMSRG using \textit{NN} N$^3$LO + 3\textit{N}(lnl) interaction.}
    \begin{tabular}{cccccccccc}
        \hline
        \hline
        \multirow{2}{*}{Nucl.} & \multicolumn{4}{c}{Parital wave} & \multirow{2}{*}{Sum} & \multicolumn{4}{c}{Contribution} \\ \cline{2-5} \cline{7-10} 
         & $0f_{7/2}$ & $1p_{3/2}$ & $0f_{5/2}$ & $1p_{1/2}$ &  & $0f_{7/2}$ & $1p_{3/2}$ & $0f_{5/2}$ & $1p_{1/2}$  \\
         \hline
         $^{48}$K & 7.29 & 1.31 & 0.06 & 0.03 & 8.70 & 84\% & 15\% & 0.7\% & 0.4\% \\
         $^{47}$Ar & 6.85 & 1.61 & 0.12 & 0.05 & 8.64 & 79\% & 19\% & 1.4\% & 0.6\% \\
         $^{46}$Cl & 5.43 & 1.62 & 0.10 & 0.11 & 7.26 & 75\% & 22\% & 1.4\% & 1.6\% \\
         $^{45}$S & 5.72 & 1.75 & 0.19 & 0.24 & 7.90 & 72\% & 22\% & 2.4\% & 3.0\% \\
         \hline
         \hline
    \end{tabular}
\label{tab:S_spectroscopic_summary}
\end{table}

Experimental data on atomic masses, excitation energies, and reduced transition probabilities have revealed that the $N=28$ shell closure weakens as protons are removed from $Z=20$~\cite{SORLIN2008602}. SFs provide direct insight into this shell evolution. Building on our previous analysis of neutron-adding processes from $^{47}$K ($^{46}$Ar) to excited states in $^{48}$K ($^{47}$Ar), we examine neutron SFs for transitions from $N=29$ nuclei to their corresponding $N=28$ isotones with various excited states, as shown in Fig.~\ref{fig:47K46Ar45Cl}, focusing on the $0f_{7/2}$, $0f_{5/2}$, $1p_{3/2}$, and $1p_{1/2}$ partial waves, the results for K and Ar serve as extensions and complements to the preceding discussion. Table~\ref{tab:S_spectroscopic_summary} lists the summed SFs for each orbital (involving the first 120 states of the $N=28$ nuclei), the total SF summed over all orbitals, and the contribution of each orbital to the total. Here, the SFs are presented without applying the reduction factor, as they directly reflect the corresponding occupation numbers and thereby quantify shell evolution. Moreover, the contributions remain unaffected in the absence of the reduction factor.
Results from VS-IMSRG calculations show that, from $^{48}$K to $^{46}$Cl, the occupation of the neutron $0f_{7/2}$ orbital decreases with decreasing proton number, while that of the neutron $1p_{3/2}$ orbital increases, implying a shrinking single-particle energy spacing between the two orbitals. This behavior is consistent with both theoretical and experimental investigations~\cite{PhysRevLett.122.222501,PhysRevLett.109.052501}, which indicate a marked erosion of the $N=28$ shell gap in nuclei below calcium.
%We further computed the effective single-particle energies (ESPEs) for the $N=28$ isotones using the VS-IMSRG with the $\textit{NN}$ N$^3$LO + 3$\textit{N}$(lnl) interaction. The resulting neutron gaps of $\nu 0f_{7/2}-\nu 1p_{3/2}$, $\nu 1p_{3/2}-\nu 1p_{1/2}$, and $\nu 1p_{1/2}-\nu 0f_{5/2}$ are $-5.61$, $-2.08$, and $-2.82$ MeV in $^{47}$K, and $-2.70$, $-1.98$, and $-3.59$ MeV in $^{44}$S, respectively. 
We further calculated the effective single-particle energies (ESPEs) for the $N=28$ isotones $^{47}$K and $^{44}$S using the VS-IMSRG approach with the $\textit{NN}$ N$^3$LO + 3$\textit{N}$(lnl) and EM1.8/2.0 interactions, as well as the LSSM calculations with the SDPF-MU and SDPF-U interactions. The results are summarized in Table~\ref{tab:ESPEs2}. In this work, the ESPE is defined as~\cite{PhysRevLett.87.082502, PhysRevC.100.034324}: 
\begin{equation}
    \varepsilon_i = \varepsilon_i^{\text core} + \sum_j n_j V_{ij}^m, 
\end{equation}
where $\varepsilon_i^{\text core}$ denotes the valence-space single-particle energy, $V^m$ is the monopole interaction, and $n_j$ is the occupation number, calculated consistently using the adopted Hamiltonian.
For all interactions considered, the neutron ESPE gap between $\nu 0f_{7/2}$ and $\nu 1p_{3/2}$ decreases as the proton number decreases. This behavior indicates a weakening of the $N=28$ shell closure when going from $^{47}$K to $^{44}$S, in agreement with the trends summarized in Table~\ref{tab:S_spectroscopic_summary}.

% %这里添加一个表格
% \begin{table*}[]
%     \centering
%     \renewcommand{\arraystretch}{1.5}
%     \caption{The ESPEs for the $N=28$ isotones using the VS-IMSRG with the $\textit{NN}$ N$^3$LO + 3$\textit{N}$(lnl) and the EM1.8/2.0 interactions, and using the LSSM with the SDPF-MU and the SDPF-U interactions. }
%     \begin{tabular}{ccccccccc}
%         \hline
%         \hline
%         & \multicolumn{4}{c}{$^{47}$K} & \multicolumn{4}{c}{$^{44}$S} \\ \cline{2-5} \cline{6-9} 
%         & $NN+3N$(lnl) & EM1.8/2.0 & SDPF-MU & SDPF-U & $NN+3N$(lnl) & EM1.8/2.0 & SDPF-MU & SDPF-U \\
%          \hline
%         $0d_{5/2}$ & 1.31 & 0.06 & 0.03 & 8.70 & 84\% & 15\% & 0.7\% & 0.4\% \\
%         $0d_{3/2}$ & 1.31 & 0.06 & 0.03 & 8.70 & 84\% & 15\% & 0.7\% & 0.4\% \\
%         $1s_{1/2}$ & 1.31 & 0.06 & 0.03 & 8.70 & 84\% & 15\% & 0.7\% & 0.4\% \\
%         $0f_{7/2}$ & 1.31 & 0.06 & 0.03 & 8.70 & 84\% & 15\% & 0.7\% & 0.4\% \\
%         $0f_{5/2}$ & 1.61 & 0.12 & 0.05 & 8.64 & 79\% & 19\% & 1.4\% & 0.6\% \\
%         $1p_{3/2}$ & 1.62 & 0.10 & 0.11 & 7.26 & 75\% & 22\% & 1.4\% & 1.6\% \\
%         $1p_{1/2}$ & 1.75 & 0.19 & 0.24 & 7.90 & 72\% & 22\% & 2.4\% & 3.0\% \\
%          \hline
%          \hline
%     \end{tabular}
% \label{tab:ESPEs1}
% \end{table*}

\begin{table*}[]
    \centering
    \renewcommand{\arraystretch}{1.5}
    \caption{The ESPEs for the $N=28$ isotones using the VS-IMSRG with the $\textit{NN}$ N$^3$LO + 3$\textit{N}$(lnl) and the EM1.8/2.0 interactions, and using the LSSM with the SDPF-MU and the SDPF-U interactions.}
    \begin{tabular}{ccccccccccccccccccccccc}
        \hline
        \hline
        & \multicolumn{2}{c}{$\pi0d_{5/2}$} && \multicolumn{2}{c}{$\pi0d_{3/2}$} && \multicolumn{2}{c}{$\pi1s_{1/2}$} && \multicolumn{2}{c}{$\nu0f_{7/2}$} && \multicolumn{2}{c}{$\nu0f_{5/2}$} && \multicolumn{2}{c}{$\nu1p_{3/2}$} && \multicolumn{2}{c}{$\nu1p_{1/2}$} \\ 
        \cline{2-3} \cline{5-6} \cline{8-9} \cline{11-12} \cline{14-15} \cline{17-18} \cline{20-21}
        & $^{47}$K & $^{44}$S & & $^{47}$K & $^{44}$S & & $^{47}$K & $^{44}$S & & $^{47}$K & $^{44}$S & & $^{47}$K & $^{44}$S & & $^{47}$K & $^{44}$S & & $^{47}$K & $^{44}$S \\
         \hline
         \textit{NN} + 3\textit{N}(lnl) & $-26.6$ & $-27.2$ & & $-19.4$ & $-19.4$ & & $-19.7$ & $-21.4$ & & $-9.88$ & $-5.67$ & & $0.63$ & $2.66$ & & $-4.27$ & $-3.02$ & & $-2.19$ & $-1.07$ \\
         EM1.8/2.0 & $-25.4$ & $-25.9$ & & $-19.4$ & $-19.7$ & & $-18.7$ & $-20.2$ & & $-9.63$ & $-5.90$ & & $-0.35$ & $1.70$ & & $-4.37$ & $-3.01$ & & $-2.58$ & $-1.35$ \\
         SDPF-MU & $-33.2$ & $-32.4$ & & $-27.7$ & $-26.0$ & & $-27.4$ & $-27.2$ & & $-11.8$ & $-7.56$ & & $-2.78$ & $0.03$ & & $-6.38$ & $-4.53$ & & $-4.41$ & $-2.77$ \\
         SDPF-U & $-42.1$ & $-40.9$ & & $-37.0$ & $-35.4$ & & $-27.4$ & $-34.4$ & & $-18.1$ & $-12.9$ & & $-8.22$ & $-4.20$ & & $-11.7$ & $-8.63$ & & $-9.36$ & $-6.22$ \\
         \hline
         \hline
    \end{tabular}
\label{tab:ESPEs2}
\end{table*}

For neutron-rich S isotopes, experimental studies have confirmed shape coexistence in $^{43}$S~\cite{PhysRevLett.102.092501, PhysRevLett.108.162501, PhysRevLett.125.232501, PhysRevLett.121.012501}, and theoretical investigations predict a similar phenomenon in $^{45}$S, indicating a reduction of the $N=28$ shell gap~\cite{PhysRevC.109.L041301}. In this work, we also calculated neutron SFs for $^{45}$S, as summarized in Table~\ref{tab:S_spectroscopic_summary}. The results reveal a pronounced reduction in the occupation of the $0f_{7/2}$ orbital, providing further evidence for the erosion of the $N=28$ magic number.

For unstable nuclei such as $^{48}$K, $^{47}$Ar, $^{46}$Cl, and $^{45}$S, an inverse kinematics experimental setup can be employed, in which a radioactive ion beam impinges on a deuteron or proton target. Neutron removal via ($d,t$) or ($p,d$) transfer reactions populates various excited states in the residual nucleus, allowing reconstruction of the neutron SF distribution. In practical experiments, only low-lying states are measurable. Our calculations indicate that transfer reactions populating states in the residual nucleus below an excitation energy of $\sim$4 MeV can be utilized to probe the strength of the $N=28$ shell closure.

%{\color{magenta}[Table 2]. Table~\ref{tab:S_spectroscopic_summary} lists the single-neutron removal spectroscopic factors (SFs) in different partial waves and the related proportions, for nuclei from $^{48}$K to $^{45}$S. For all the cases, we calculate 120 states for the residual nucleus; the final summed value still deviates slightly from the expected theoretical value of 9. Despite this, the relative contributions from each single-particle orbital clearly reveal the underlying shell evolution. From $^{48}$K to $^{45}$S, the occupation of the $0f_{7/2}$ orbital decreases, while that of the $1p_{3/2}$ orbital increases. This trend indicates a reduction in the energy gap between the $0f_{7/2}$ and $1p_{3/2}$ orbitals, reflecting the gradual disappearance of the $N=28$ shell closure as the proton number decreases along the $N=29$ isotonic chain.}

\section{Summary}
%\textit{Summary.---}
We revisit the SFs and excitation energies of low-lying states in $^{48}$K, recently probed with the $^{47}$K$(d,p\gamma)^{48}$K transfer reaction, by performing \textit{ab initio} VS-IMSRG approach based on chiral $NN+$3$N$ forces.
Our VS-IMSRG calculations reproduce the level ordering for low-lying states in $^{48}$K when using the \textit{NN} N$^3$LO + 3\textit{N}(lnl) interaction. However, the calculated SFs exhibit a systematic overestimation compared to experimental data, which is similar to the LSSM results.
We resolve this discrepancy by introducing a reduction factor, commonly involved in direct reaction theory but previously omitted in analyses of $^{48}$K. Upon incorporating this factor, the \textit{ab initio} VS-IMSRG SF values, as well as those from LSSM calculations, achieve excellent agreement with the experimental data for $^{48}$K and $^{47}$Ar.
Furthermore, leveraging the \textit{ab initio} VS-IMSRG framework with the \textit{NN} N$^3$LO + 3\textit{N}(lnl) interaction, we extend our analysis to calculate the single-neutron removal SFs for the valence 
neutron across the $N=29$ isotones, ranging from $^{48}$K to $^{45}$S. The systematic trend observed in these calculated SFs provides direct evidence for the progressive weakening of the $N=28$ subshell closure in this isotopic chain. 

\textit{Acknowledge.}---
This work has been supported by the National Key R\&D Program of China under Grant No. 2023YFA1606403; the National Natural Science Foundation of China under Grant Nos. 12205340, 12175281, and 12475128; the Gansu Natural Science Foundation under Grant Nos. 22JR5RA123 and 25JRRA467; %W.J.H. acknowledges support from 
the Talent Support Project of Guangdong Program No. 2023TQ07A872; The numerical calculations in this paper have been done at Hefei Advanced Computing Center.

\bibliography{ref}

\end{document}